# Coherent Spin-Phonon Coupling in the Layered Ferrimagnet Mn$_3$Si$_2$Te$_6$

L. M. Martinez[1,2], Y. Liu[3,¶], C. Petrovic[3], S. Haldar[4], T. Griepe[5], U. Atxitia[5],

M. Campbell[1,6], M. Pettes[1], R. P. Prasankumar[1],

E. J. G. Santos[4,7]*, S. R. Singamaneni[2,‡], and P. Padmanabhan[1,†]

[1]Center for Integrated Nanotechnologies, Los Alamos National Laboratory, Los Alamos, NM 87545

[2]The University of Texas at El Paso, Department of Physics, El Paso, TX 79968

[3]Brookhaven National Laboratory, Exploratory Materials Synthesis & Characterization, Upton, NY 11973

[4]Institute for Condensed Matter Physics and Complex Systems, School of Physics and Astronomy, University of Edinburgh, Edinburgh EH9 3FD, United Kingdom

[5]Instituto de Ciencia de Materiales de Madrid, CSIC, Cantoblanco, 28049 Madrid, Spain

[6]University of California Irvine, Department of Physics & Astronomy, Irvine, CA 92697

[7]Higgs Centre for Theoretical Physics, The University of Edinburgh, EH9 3FD, United Kingdom

†prashpad@lanl.gov

*esantos@ed.ac.uk

‡srao@utep.edu

¶Present address: Los Alamos National Laboratory, Los Alamos, New Mexico 87545, USA

**Abstract**

We utilize ultrafast photoexcitation to drive coherent lattice oscillations in the layered ferrimagnetic crystal Mn$_3$Si$_2$Te$_6$, which significantly stiffen below the magnetic ordering temperature. We suggest that this is due to an exchange-mediated contraction of the lattice, stemming from strong magneto-structural coupling in this material. Additionally, simulations of the transient incoherent dynamics reveal the importance of spin relaxation channels mediated by optical and acoustic phonon scattering. Our findings highlight the importance of spin-lattice coupling in van der Waals magnets and a promising route for their dynamic optical control through their intertwined electronic, lattice, and spin degrees of freedom.

Ultrafast optical spectroscopy has been extensively used to study the dynamics of a variety of novel phenomena, such as the coherent optical excitation of phonons in solids [1–4], the driving of metastable phase transitions [5–8], and the dynamic control of magnetic materials [9–11]. In particular, the optical control of magnetic order is of great importance to push the bounds of magnetic memory storage [12] and spintronic [13,14] devices. However, in order to achieve this, a clear understanding of how the charge, spin, and lattice subsystems interact on their intrinsic timescales is of critical importance [4,15–19]. With the recent discovery of van der Waals (vdW) materials hosting intrinsic magnetic order down to the monolayer limit, there has been a new push to explore how ultrafast photoexcitation can be exploited to manipulate spin, structural order, and related properties at the nanoscale [20–22]. This, in turn, could yield revolutionary advances relevant to nanoscale information storage [9,23], spintronic applications [24–26], and proximal control of complementary nanomaterials [9,23].

In addition to thickness (i.e., layer number), the properties of vdW magnets are highly sensitive to external stimuli [27], making them ideal candidate materials from the perspective of dynamically reconfigurable nanoscale magnetic devices. Various studies have shown how their magnetic properties can be altered through pressure, chemical doping, proton irradiation, and photoexcitation [28–34]. The latter is particularly interesting because it is a contactless route for high-speed control of magnetic states. Recently, intense femtosecond pulses have proven successful in driving picosecond de- and re-magnetization [4,35], all-optical switching of magnetic states in an atomically thin ferromagnet [36], the coherent excitation of GHz magnons [37], and revealing the presence of spin-coupled coherent lattice vibrations [4]. The latter is due to strong coupling between the lattice and spin degrees of freedom, a feature that has proven to be of importance in a variety of layered magnetic materials.

In this letter, we explore the electronic, spin, and structural dynamics of the psuedo-vdW ferrimagnet $Mn_3Si_2Te_6$ driven by femtosecond optical pulses. By applying a modified microscopic four temperature model (M4TM) [38], we are able to extract the electron-phonon coupling constants and simulate the electronic dynamics. This reveals a robust spin relaxation channel through optical and acoustic phonon scattering. Moreover, we reveal the presence of strong coherent lattice vibrations whose oscillatory frequency is highly sensitive to the onset of ferrimagnetic order. We observe a pronounced temperature-dependent stiffening of the phonon below the critical temperature ($T_c$) that coincides with the temperature-dependent magnetization. Such order-driven dynamic renormalization of *coherent* vibrational frequencies is, to the best of our knowledge, yet to be observed in vdW magnetic materials and provides compelling support for previous assertions [39,40] of the strength and robustness of spin-lattice coupling in $Mn_3Si_2Te_6$ and related vdW magnetic materials.

As shown in Fig. 1(a), $Mn_3Si_2Te_6$ is a self-intercalated layered magnetic material comprised of two distinct magnetic sublattices. The first sublattice hosts an Mn site ($Mn^1$) which is within a hexagonal structure, typical of vdW materials. The second sublattice contains another Mn site ($Mn^2$) which fills a third of the octahedral holes within the vdW gap [39,41,42]. The $Mn^1$ moments lie primarily along the *ab*-axis while the $Mn^2$ moments are aligned in an antiparallel fashion [41,43]. This magnetic structure gives rise to frustration [41], setting the stage for ferrimagnetic order below $T_c \approx 75$ K [41,44]. The unusual magnetic order also gives rise to nodal line degeneracies, which when lifted with external magnetic fields, reveal the presence of colossal magnetoresistance [43] due to helically aligned chiral orbital currents that circulate along the Te-Te edges of the Mn-Te octahedra [40]. Additionally, this field also drives a concomitant reduction in the $ab$-plane lattice parameter, pointing to the presence of strong spin-lattice coupling. Recently, spontaneous Raman studies have characterized the temperature evolution of various phonon modes in Mn$_3$Si$_2$Te$_6$. Djurdjić *et al.* found two Raman modes with A$_{1g}$ symmetry that provide evidence of three successive phase transitions (between 142 - 285 K) that could arise from competing magnetic states known to exist in the material [45]. Nevertheless, the behavior of the Raman modes below $T_c$ has yet to be quantified. To date, the manifestation of the strong spin-lattice coupling, alluded to in Ref. [39], particularly under dynamic non-equilibrium conditions, has yet to be investigated.

In our study, bulk-like flakes of single crystal Mn$_3$Si$_2$Te$_6$ were prepared via the self-flux method [39,42]. The sample was mechanically exfoliated and transferred to an optical cryostat in an argon atmosphere to prevent the formation of TeO$_2$ complexes [45]. Ultrafast transient reflectivity measurements were then performed using a regeneratively amplified Ti:sapphire laser system with a repetition rate of 100 kHz. The pump and probe photon energies were 1.55 eV with a nominal pulse width of ~60 femtoseconds (fs). The photon energy used in this work is higher than the estimated bandgap (~0.2 eV) for this material [43,46]. The laser beam was split into pump and probe arms that were focused onto the sample at near-normal incidence using a 20X apochromatic objective (to ~7 μm diameters) and linearly cross-polarized in order to suppress spurious signals due to pump scattering. The probe fluence was held at 0.125 mJ/cm$^2$ while the pump fluence was varied with a maximal limit of 2.5 mJ/cm$^2$ to mitigate sample damage. The pump and probe beams were separately modulated at $f_{\text{pump}} = 0.7$ kHz and $f_{\text{probe}} = 0.5$ kHz, and the pump-induced changes in the probe reflectivity were measured using phase-sensitive detection at the $f_{\text{pump}} + f_{\text{probe}}$ intermodulation frequency.

Figure 1(b) shows the differential reflectivity ($\Delta R/R$) as a function of temperature. Two main components are observed, namely a broad time-dependent incoherent background superimposed with pronounced coherent oscillations. The former is due to the nonequilibrium dynamics associated with hot electrons driven by the above bandgap photoexcitation. This can be fit using a biexponential function [2,18],

$$\frac{\Delta R}{R} = Ae^{-\frac{t}{\tau_1}} + Be^{-\frac{t}{\tau_2}} \tag{1}$$

where the amplitudes ($A$ and $B$) and time constants ($\tau_1$ and $\tau_2$) are treated as free parameters. The black curves in Fig. 1(b) show the best fits to the temperature-dependent data, revealing a relatively fast rise followed by a slow relaxation, which is consistent across the entire temperature range. Similar fits were calculated for the pump-fluence-dependent measurements [Figure 1(c)], which readily reveal an increase in the magnitude of $A$ and $B$ at higher fluences [see Supp. Mat. Fig. 1], consistent with their hot electron origin.

To delve deeper into the incoherent dynamics, we developed a M4TM approach [38]. This model allows us to describe the coupling of hot electrons to the lattice system through both the optical and acoustic phonon subsystems, in addition to the spin system. The system is defined by the temperatures of the electronic, optical phonon, and acoustic phonon temperatures ($T_e, T_{po}$, and $T_{pa}$, respectively) and their heat capacities ($C_e, C_{po}$, and $C_{pa}$, respectively). Splitting the lattice system into optical and acoustic phonons is an approximation due to acoustic phonons also undergoing electron-phonon scattering, but at a reduced rate. The electron and two phonon dynamics are described by the coupled equations:

$$C_e \frac{dT_e}{dt} = g_{e-po}(T_{po} - T_e) + P(t) + \dot{Q}_{e-s}, \tag{2}$$

$$C_{po} \frac{dT_{po}}{dt} = g_{e-po}(T_{po} - T_e) + g_{po-pa}(T_{pa} - T_{po}), \tag{3}$$

$$C_{pa} \frac{dT_{pa}}{dt} = g_{po-pa}(T_{pa} - T_{po}), \tag{4}$$

where $P(t)$ is the absorbed laser pulse energy that excites the electronic system, $g_{e-po}$ is the coupling between electron and optical phonons, and $g_{po-pa}$ is the optical-phonon-acoustic-phonon coupling constant. The electron system is energetically coupled to (i) the optical phonons with temperature $T_{po}$ and (ii) the spin system via the energy flow, $\dot{Q}_{e-s} = \frac{Jm\dot{m}}{V_{at}}$, where $J$ is defined by $J = 3\frac{S^2}{S(S+1)}k_B T_c$, where $m$ is the magnetization, $S$ is the spin, and $V_{at}$ is the atomic volume calculated by dividing the unit cell volume by the number of atoms in a unit cell [47]. The optical phonons are also coupled to the acoustic phonons with temperature $T_{pa}$ and heat capacity $C_{pa}$. Additional details of the M4TM description are shown in the Supplemental Material. Typically, this phenomenological description of the microscopic interactions between the electrons, phonons, and spins is used to describe the dynamics in metallic systems [38]. However, this model has been recently applied to semiconducting systems due to the transient quasi-metallic state driven by the ultrafast photoexcitation of carriers [4,38]. Within this model, the spin dynamics are not internally thermalized and can be described by the electron origin of the spin-flip process [see Supp. Mat.].

The use of the M4TM allows us to extract the temperature dependences of key parameters in the system, such as $g_{e-po}$, the Sommerfield coefficient ($\gamma_e$), $C_e$, $C_{po}$, $C_{pa}$, and the probability of phonon-mediated spin-flip scattering ($a_{sf}$). The latter, enabled by strong spin-orbit coupling leading to the majority and minority carrier spins being admixtures of pure spin states [48], has been suggested to be a major driver of ultrafast demagnetization in prototypical vdW magnets [4]. The model parameters obtained for $Mn_3Si_2Te_6$ between 25 - 40 K are shown in Table I. In addition, Figure 2(a) shows the M4TM simulation of the incoherent dynamics at 25 K, demonstrating good qualitative agreement with the experimental data (additional M4TM simulations of the experimental data up to 75 K are shown in Supp. Mat. Fig. 3). We find that the M4TM provides a consistent model of the incoherent dynamics driven by the energy flow between the electrons, phonons, and spin subsystems up to 75 K. $g_{e-po}$ is estimated to be approximately $\sim 1.85 \times 10^{18}$ W/m³K, showing minimal variation with temperature, and is significantly larger than the recently reported electron-phonon coupling for a similar vdW magnet $Cr_2Ge_2Te_6$ ($\sim 15 \times 10^{16}$ W/m³K) [38]. More importantly, $a_{sf}$ steadily increases below $T_c$, and its relatively large value, in line with $CrI_3$ [4], supports the presence of strong spin-phonon coupling in $Mn_3Si_2Te_6$.

More insight can be gleaned from the biexponential fits obtained from Eq. 1. $\tau_1$ has values on the order of ~350 fs that increase in a linear fashion below $T_c$ [blue dots in Fig. 2(b)]. Because our pump energy far exceeds the estimated bandgap [43,46], the carrier dynamics on this time scale are most likely associated with electron-electron scattering, which is suppressed at lower temperatures. Other magnetic semiconductors have shown similar temperature variations in the sub-picosecond carrier dynamics [2,18,49,50]. $\tau_2$, associated with electron-phonon scattering processes, increases slightly from 125 to 60 K [blue dots in Fig. 2(c)]. This trend reverses below 60 K, where the electron-phonon scattering becomes faster, reaching a value of 13 ps at 25 K. This is likely due to the enhanced spin-phonon coupling below $T_c$ and the activation of additional phonon-mediated spin-flip scattering channels, an assertion supported by the mirrored temperature dependence of $a_{sf}$. The time constants extracted from the M4TM are presented (black dots) in Fig. 2(b) and 2(c). These show a good qualitative agreement with the experimental values, differentiated only by a fixed offset. We note that data above 70 K are not reproduced well by the M4TM due to the lack of any spin ordering above the magnetic phase transition.

We now turn our attention to the coherent oscillatory dynamics. Figure 3(a) shows the measured residual $\Delta R/R$ ($\Delta R_{res}/R$) obtained by subtracting out the electronic background discussed above. Strong coherent oscillations are observed with a characteristic decay time of ~3 ps and an oscillatory frequency of ~3.52 THz. In Fig. 3(b), we plot the Fourier transform (FT) of the time-domain data for various temperatures. Based on recent spontaneous Raman scattering measurements [45], we identify this mode as a phonon of $E_g$ symmetry, driven via an impulsive stimulated Raman scattering process by the ultrafast pump [51,52]. We observe a subtle, yet significant, stiffening of the mode at lower temperatures, evidenced by the shift of the FT peak to higher frequencies in Fig. 3(b). To better capture the temperature-dependent variation of the phonon frequency, we fit a damped sine function to the $\Delta R_{res}/R$ [51],

$$\frac{\Delta R_{res}}{R} = A_{DO} e^{-\frac{t}{\tau_{DO}}} \sin(\omega_{DO} t + \phi_{DO}) \quad (5)$$

where $A_{DO}$ is the amplitude, $\tau_{DO}$ is the time constant, $\omega_{DO}$ is the frequency, and $\phi_{DO}$ is the phase shift of the damped oscillator. The black lines in Fig. 3(a) are the fits using Eq. 5. We plot the temperature dependence of $\omega_{DO}$ in Fig. 4. The frequency is relatively constant (~3.53 THz) above $T_c$, but increases rapidly from 75 K down to 50 K, before stabilizing to ~3.57 THz at lower temperatures. This temperature dependence is distinct from what would be expected under the assumption of thermally driven lattice contraction [53]. This can be modeled with the temperature-dependent phonon frequency, $\omega(T)$, with contributions from both anharmonic decay and thermal lattice contraction [53], and primarily dependent on the thermal occupation of the phonons. This is expressed as

$$\omega(T) = \omega_o \left[ 1 - \gamma \frac{V(T) - V_o}{V_o} - \left(\frac{\Gamma_{L,0}}{\sqrt{2}\omega_o}\right)^2 \left(1 + \frac{4\lambda_{ph-ph}}{\exp\left(\frac{\omega_o}{2k_B T - 1}\right)}\right) \right]. \quad (6)$$

Here, $V(T)$ and $V_o$ are the changes in the unit cell volume with varying temperature $T$ and $T \to 0$ obtained from Ref. [41], respectively. $\gamma$, $\Gamma_{L,0}$, $\omega_0$, and $\lambda_{ph-ph}$ are free parameters in the fitting of Eq. 6 to $\omega_{DO}(T)$, with $\gamma$ being the Grüneisen parameter, $\Gamma_{L,0}$ and $\omega_0$ being the linewidth and frequency of the phonon near the zero temperature limits, respectively, and $\lambda_{ph-ph}$ being the phonon-phonon coupling. As seen from the red dashed line in Fig. 4, the thermal expansion model fails to reproduce the sigmoidal shape of the experimental $\omega_{DO}(T)$.

We have previously shown that magnetic exchange interactions can be modulated by distortions of the structural lattice in vdW magnets [4], leading to spin-coupled coherent phonon dynamics. Here, we postulate the inverse of this effect, where the onset of ferrimagnetic order leads to dynamic exchange magnetostriction [54], yielding a spin-driven contraction of the lattice. An overlay of the temperature-dependent magnetization (blue curve in Fig. 4), obtained via SQUID measurements, reveals a relatively close agreement with the experimental $\omega_{D0}(T)$ (given the laser-indued heating of the lattice). This indicates that the temperature dependence of the phonon frequency roughly follows the onset of ferrimagnetic order in the crystal, underscoring its spin-driven origin. Indeed, similar spin-order dependent shifts of *thermal* phonon frequencies have been observed in other materials [55–57] through spontaneous Raman scattering. However, a spin-order dependent shift of the *coherent* phonon frequency has, to the best of our knowledge, never been observed in any vdW magnet. We note that, like CrI$_3$ [4], the coherent phonon mode in Mn$_3$Si$_2$Te$_6$ should also demonstrate a spin-coupled character, which can be verified using techniques such as time-resolved magneto-optical Kerr effect spectroscopy in future studies.

In conclusion, we use ultrafast optical spectroscopy to study electronic and coherent phonon dynamics in the ferrimagnetic crystal Mn$_3$Si$_2$Te$_6$. By using the M4TM, we are able to model the electronic dynamics due to the presence of the strong spin-lattice coupling below 70 K. Most importantly, we identify a coherent phonon mode that couples strongly to ferrimagnetic order. This phonon's frequency shift across $T_c$ mirrors the temperature-dependent magnetization and most likely originates from an exchange mediated interaction between the coherent phonon and spins. Additional studies focusing on the spin dynamics could provide greater insight into the relationship between the phonon and spin systems in Mn$_3$Si$_2$Te$_6$. This is of crucial interest in the evolving field of vdW magnetism, as we look to uncover the subtle yet profound connection between lattice structure and spin-driven properties in these materials, which have critical applications in emerging nanoelectronic devices. For example, dynamic optical control over spin-coupled lattice modes could open new routes to harnessing the magnetoresistive properties of Mn$_3$Si$_2$Te$_6$ for high-speed reconfigurable conductive devices needed for advanced computing architectures.

**Acknowledgements**

P.P. acknowledges support from the Los Alamos National Laboratory (LANL) Laboratory Directed Research and Development (LDRD) program. L.M.M. and S.R.S. acknowledge support from the National Science Foundation (NSF) – Division of Materials Research (Award No. 21051091). S.R.S. further acknowledges support from NSF – Major Research Instrumentation program (Award No. 2018067). Work at Brookhaven National Laboratory is supported by the U.S. Department of Energy (DOE), Office of Science, Basic Energy Sciences, Materials Sciences and Engineering Division under Contract No. DESC0012704 (materials synthesis). E.J.G.S. acknowledges computational resources through CIRRUS Tier-2 HPC Service (ec131 Cirrus Project) at EPCC (http://www.cirrus.ac.uk) funded by the University of Edinburgh and Engineering and Physical Sciences Research Council (EP/P020267/1); ARCHER UK National Supercomputing Service (http://www.archer.ac.uk) via Project d429. E.J.G.S. further acknowledges the EPSRC Open Fellowship (EP/T021578/1) and the Edinburgh–Rice Strategic

Collaboration Awards for funding support. This work was performed, in part, at the Center for Integrated Nanotechnologies, an Office of Science User Facility operated for the U.S. DOE Office of Science, under user proposal #2022AU0117. LANL, an affirmative action equal opportunity employer, is managed by Triad National Security, LLC for the U.S. DOE's National Nuclear Security Administration, under contract 89233218CNA000001.

**References**

[1] M.-C. Lee et al., *Strong Spin-Phonon Coupling Unveiled by Coherent Phonon Oscillations in $Ca_2RuO_4$*, Phys. Rev. B **99**, 144306 (2019).

[2] J. Guo, W. Liang, and S.-N. Luo, *Anomalous Hot Carrier Decay in Ferromagnetic $Cr_2Ge_2Te_6$ via Spin–Phonon Coupling*, J. Phys. Chem. Lett. **11**, 9351 (2020).

[3] K. J. Yee, K. G. Lee, E. Oh, D. S. Kim, and Y. S. Lim, *Coherent Optical Phonon Oscillations in Bulk GaN Excited by Far below the Band Gap Photons*, Phys. Rev. Lett. **88**, 105501 (2002).

[4] P. Padmanabhan et al., *Coherent Helicity-Dependent Spin-Phonon Oscillations in the Ferromagnetic van Der Waals Crystal $CrI_3$*, Nat. Commun. **13**, 1 (2022).

[5] M. Fiebig, K. Miyano, Y. Tomioka, and Y. Tokura, *Visualization of the Local Insulator-Metal Transition in $Pr_{0.7}Ca_{0.3}MnO_3$*, Science **280**, 1925 (1998).

[6] M. Matsubara, Y. Okimoto, T. Ogasawara, Y. Tomioka, H. Okamoto, and Y. Tokura, *Ultrafast Photoinduced Insulator-Ferromagnet Transition in the Perovskite Manganite $Gd_{0.55}Sr_{0.45}MnO_3$*, Phys. Rev. Lett. **99**, 207401 (2007).

[7] C. A. Belvin et al., *Exciton-Driven Antiferromagnetic Metal in a Correlated van Der Waals Insulator*, Nat. Commun. **12**, 1 (2021).

[8] A. S. Disa et al., *Photo-Induced High-Temperature Ferromagnetism in $YTiO_3$*, Nature **617**, 7959 (2023).

[9] A. V. Kimel, A. Kirilyuk, and T. Rasing, *Femtosecond Opto-Magnetism: Ultrafast Laser Manipulation of Magnetic Materials*, Laser Photonics Rev. **1**, 275 (2007).

[10] A. Kirilyuk, A. V. Kimel, and T. Rasing, *Ultrafast Optical Manipulation of Magnetic Order*, Rev. Mod. Phys. **82**, 3 (2010).

[11] N. Wu, S. Zhang, Y. Wang, and S. Meng, *Ultrafast All-Optical Quantum Control of Magnetization Dynamics*, Prog. Surf. Sci. 100709 (2023).

[12] A. V. Kimel and M. Li, *Writing Magnetic Memory with Ultrashort Light Pulses*, Nat. Rev. Mater. **4**, 3 (2019).

[13] J. Pettine, P. Padmanabhan, N. Sirica, R. P. Prasankumar, A. J. Taylor, and H.-T. Chen, *Ultrafast Terahertz Emission from Emerging Symmetry-Broken Materials*, Light Sci. Appl. **12**, 1 (2023).

[14] E. T. Papaioannou and R. Beigang, *THz Spintronic Emitters: A Review on Achievements and Future Challenges*, Nanophotonics **10**, 1243 (2021).

[15] D. Afanasiev, J. R. Hortensius, B. A. Ivanov, A. Sasani, E. Bousquet, Y. M. Blanter, R. V. Mikhaylovskiy, A. V. Kimel, and A. D. Caviglia, *Ultrafast Control of Magnetic Interactions via Light-Driven Phonons*, Nat. Mater. **20**, 5 (2021).

[16] M. Hase, K. Mizoguchi, H. Harima, S. Nakashima, M. Tani, K. Sakai, and M. Hangyo, *Optical Control of Coherent Optical Phonons in Bismuth Films*, Appl. Phys. Lett. **69**, 2474 (1996).

[17] M. Först, C. Manzoni, S. Kaiser, Y. Tomioka, Y. Tokura, R. Merlin, and A. Cavalleri, *Nonlinear Phononics as an Ultrafast Route to Lattice Control*, Nat. Phys. **7**, 11 (2011).

[18] D. J. Lovinger et al., *Magnetoelastic Coupling to Coherent Acoustic Phonon Modes in the Ferrimagnetic Insulator $GdTiO_3$*, Phys. Rev. B **102**, 085138 (2020).

[19] H. Ling and A. R. Davoyan, *Light Control with Atomically Thin Magnets*, Nat. Photonics **16**, 4 (2022).

[20] M. Khela, M. Dąbrowski, S. Khan, P. S. Keatley, I. Verzhbitskiy, G. Eda, R. J. Hicken, H. Kurebayashi, and E. J. G. Santos, *Laser-Induced Topological Spin Switching in a 2D van Der Waals Magnet*, Nat. Commun. **14**, 1 (2023).

[21] M. Dąbrowski, S. Guo, M. Strungaru, P. S. Keatley, F. Withers, E. J. G. Santos, and R. J. Hicken, *All-Optical Control of Spin in a 2D van Der Waals Magnet*, Nat. Commun. **13**, 1 (2022).

[22] M. Strungaru, M. Augustin, and E. J. G. Santos, *Ultrafast Laser-Driven Topological Spin Textures on a 2D Magnet*, Npj Comput. Mater. **8**, 1 (2022).

[23] A. V. Kimel and M. Li, *Writing Magnetic Memory with Ultrashort Light Pulses*, Nat. Rev. Mater. **4**, 3 (2019).

[24] P. Němec, M. Fiebig, T. Kampfrath, and A. V. Kimel, *Antiferromagnetic Opto-Spintronics*, Nat. Phys. **14**, 3 (2018).

[25] V. P. Ningrum et al., *Recent Advances in Two-Dimensional Magnets: Physics and Devices towards Spintronic Applications*, Research **2020**, (2020).

[26] G. Wu, S. Chen, Y. Ren, Q. Y. Jin, and Z. Zhang, *Laser-Induced Magnetization Dynamics in Interlayer-Coupled $[Ni/Co]_4/Ru/[Co/Ni]_3$ Perpendicular Magnetic Films for Information Storage*, ACS Appl. Nano Mater. **2**, 5140 (2019).

[27] S. Yang, T. Zhang, and C. Jiang, *Van Der Waals Magnets: Material Family, Detection and Modulation of Magnetism, and Perspective in Spintronics*, Adv. Sci. **8**, 2002488 (2021).

[28] B. Liu et al., *Light-Tunable Ferromagnetism in Atomically Thin $Fe_3GeTe_2$ Driven by Femtosecond Laser Pulse*, Phys. Rev. Lett. **125**, 267205 (2020).

[29] S. Jiang, L. Li, Z. Wang, K. F. Mak, and J. Shan, *Controlling Magnetism in 2D $CrI_3$ by Electrostatic Doping*, Nat. Nanotechnol. **13**, 549 (2018).

[30] K. S. Burch, *Electric Switching of Magnetism in 2D*, Nat. Nanotechnol. **13**, 532 (2018).

[31] M. Abramchuk, S. Jaszewski, K. R. Metz, G. B. Osterhoudt, Y. Wang, K. S. Burch, and F. Tafti, *Controlling Magnetic and Optical Properties of the van Der Waals Crystal $CrCl_{3−x}Br_x$ via Mixed Halide Chemistry*, Adv. Mater. **30**, 1801325 (2018).

[32] S. Mondal, M. Kannan, M. Das, L. Govindaraj, R. Singha, B. Satpati, S. Arumugam, and P. Mandal, *Effect of Hydrostatic Pressure on Ferromagnetism in Two-Dimensional $CrI_3$*, Phys. Rev. B **99**, 180407 (2019).

[33] L. M. Martinez, H. Iturriaga, R. Olmos, L. Shao, Y. Liu, T. T. Mai, C. Petrovic, A. R. Hight Walker, and S. R. Singamaneni, *Enhanced Magnetization in Proton Irradiated $Mn_3Si_2Te_6$ van Der Waals Crystals*, Appl. Phys. Lett. **116**, 172404 (2020).

[34] Q. H. Wang et al., *The Magnetic Genome of Two-Dimensional van Der Waals Materials*, ACS Nano **16**, 6960 (2022).

[35] T. Sun et al., *Ultra-Long Spin Relaxation in Two-Dimensional Ferromagnet $Cr_2Ge_2Te_6$ Flake*, 2D Mater. **8**, 045040 (2021).

[36] P. Zhang, T.-F. Chung, Q. Li, S. Wang, Q. Wang, W. L. B. Huey, S. Yang, J. E. Goldberger, J. Yao, and X. Zhang, *All-Optical Switching of Magnetization in Atomically Thin $CrI_3$*, Nat. Mater. **21**, 12 (2022).

[37] X.-X. Zhang, L. Li, D. Weber, J. Goldberger, K. F. Mak, and J. Shan, *Gate-Tunable Spin Waves in Antiferromagnetic Atomic Bilayers*, Nat. Mater. **19**, 8 (2020).

[38] E. Sutcliffe, X. Sun, I. Verzhbitskiy, T. Griepe, U. Atxitia, G. Eda, E. J. G. Santos, and J. O. Johansson, *Transient Magneto-Optical Spectrum of Photoexcited Electrons in the van Der Waals Ferromagnet $Cr_2Ge_2Te_6$*, Phys. Rev. B **107**, 174432 (2023).

[39] Y. Liu, Z. Hu, M. Abeykoon, E. Stavitski, K. Attenkofer, E. D. Bauer, and C. Petrovic, *Polaronic Transport and Thermoelectricity in $Mn_3Si_2Te_6$ Single Crystals*, Phys. Rev. B **103**, 245122 (2021).

[40] Y. Zhang, Y. Ni, H. Zhao, S. Hakani, F. Ye, L. DeLong, I. Kimchi, and G. Cao, *Control of Chiral Orbital Currents in a Colossal Magnetoresistance Material*, Nature **611**, 7936 (2022).

[41] A. F. May, Y. Liu, S. Calder, D. S. Parker, T. Pandey, E. Cakmak, H. Cao, J. Yan, and M. A. McGuire, *Magnetic Order and Interactions in Ferrimagnetic $Mn_3Si_2Te_6$*, Phys. Rev. B **95**, 174440 (2017).

[42] Y. Liu and C. Petrovic, *Critical Behavior and Magnetocaloric Effect in $Mn_3Si_2Te_6$*, Phys. Rev. B **98**, 064423 (2018).

[43] J. Seo et al., *Colossal Angular Magnetoresistance in Ferrimagnetic Nodal-Line Semiconductors*, Nature **599**, 7886 (2021).

[44] F. Ye, M. Matsuda, Z. Morgan, T. Sherline, Y. Ni, H. Zhao, and G. Cao, *Magnetic Structure and Spin Fluctuations in the Colossal Magnetoresistance Ferrimagnet $Mn_3Si_2Te_6$*, Phys. Rev. B **106**, L180402 (2022).

[45] S. D. Mijin, A. Šolajić, J. Pešić, Y. Liu, C. Petrovic, M. Bockstedte, A. Bonanni, Z. V. Popović, and N. Lazarević, *Spin-Phonon Interaction and Short Range Order in $Mn_3Si_2Te_6$*, Phys. Rev. B **107**, 054309 (2023).

[46] Y. Zhang, L.-F. Lin, A. Moreo, and E. Dagotto, *Electronic Structure, Magnetic Properties, Spin Orientation, and Doping Effect in $Mn_3Si_2Te_6$*, arXiv:2211.13321.

[47] T. Griepe and U. Atxitia, *Evidence of Electron-Phonon Mediated Spin Flip as Driving Mechanism for Ultrafast Magnetization Dynamics in 3d Ferromagnets*, Phys. Rev. B **107**, L100407 (2023).

[48] K. Carva, M. Battiato, and P. M. Oppeneer, *Ab Initio Investigation of the Elliott-Yafet Electron-Phonon Mechanism in Laser-Induced Ultrafast Demagnetization*, Phys. Rev. Lett. **107**, 207201 (2011).

[49] T. Lichtenberg, C. F. Schippers, S. C. P. van Kooten, S. G. F. Evers, B. Barcones, M. H. D. Guimarães, and B. Koopmans, *Anisotropic Laser-Pulse-Induced Magnetization Dynamics in van Der Waals Magnet $Fe_3GeTe_2$*, 2D Mater. **10**, 015008 (2022).

[50] D. Bossini, S. Dal Conte, M. Terschanski, G. Springholz, A. Bonanni, K. Deltenre, F. Anders, G. S. Uhrig, G. Cerullo, and M. Cinchetti, *Femtosecond Phononic Coupling to Both Spins and Charges in a Room-Temperature Antiferromagnetic Semiconductor*, Phys. Rev. B **104**, 224424 (2021).

[51] K. Ishioka and O. V. Misochko, *Coherent Lattice Oscillations in Solids and Their Optical Control*, in *Progress in Ultrafast Intense Laser Science: Volume V*, edited by K. Yamanouchi, A. Giulietti, and K. Ledingham (Springer, Berlin, Heidelberg, 2010), pp. 23–46.

[52] T. Dekorsy, G. C. Cho, and H. Kurz, *Coherent Phonons in Condensed Media*, in *Light Scattering in Solids VIII: Fullerenes, Semiconductor Surfaces, Coherent Phonons*, edited by M. Cardona and G. Güntherodt (Springer, Berlin, Heidelberg, 2000), pp. 169–209.

[53] A. Baum et al., *Phonon Anomalies in FeS*, Phys. Rev. B **97**, 054306 (2018).
[54] S. Jiang, H. Xie, J. Shan, and K. F. Mak, *Exchange Magnetostriction in Two-Dimensional Antiferromagnets*, Nat. Mater. **19**, 12 (2020).
[55] Y. Tian, M. J. Gray, H. Ji, R. J. Cava, and K. S. Burch, *Magneto-Elastic Coupling in a Potential Ferromagnetic 2D Atomic Crystal*, 2D Mater. **3**, 025035 (2016).
[56] K. Kim, S. Y. Lim, J.-U. Lee, S. Lee, T. Y. Kim, K. Park, G. S. Jeon, C.-H. Park, J.-G. Park, and H. Cheong, *Suppression of Magnetic Ordering in XXZ-Type Antiferromagnetic Monolayer NiPS$_3$*, Nat. Commun. **10**, 1 (2019).
[57] T. Yin et al., *Chiral Phonons and Giant Magneto-Optical Effect in CrBr$_3$ 2D Magnet*, Adv. Mater. **33**, 2101618 (2021).

Table I. Parameters obtained from the M4TM used to simulate the experimental $\Delta R/R$. $g_{e-po}$ is the electron-phonon coupling constant, $\gamma_e$ is the Sommerfield coefficient, $C_{po}$ is the optical phonon heat capacity, $C_{pa}$ is the acoustic phonon heat capacity and $a_{sf}$ is the spin-flip probability.

| T (K) | $g_{e-po}$ (W/m³ K) | $\gamma_e$ (J/m³ K²) | $C_{po}$ (J/m³ K) | $C_{pa}$ (J/m³ K) | $a_{sf}$ |
|---|---|---|---|---|---|
| 25 | $1.75 \times 10^{18}$ | $4.4 \times 10^3$ | $3.2 \times 10^7$ | $3.0 \times 10^6$ | 0.180 |
| 30 | $1.85 \times 10^{18}$ | $4.0 \times 10^3$ | $4.0 \times 10^7$ | $2.0 \times 10^6$ | 0.150 |
| 35 | $1.85 \times 10^{18}$ | $4.0 \times 10^3$ | $5.3 \times 10^7$ | $1.0 \times 10^6$ | 0.125 |
| 40 | $1.95 \times 10^{18}$ | $4.4 \times 10^3$ | $7.8 \times 10^7$ | $1.0 \times 10^6$ | 0.115 |

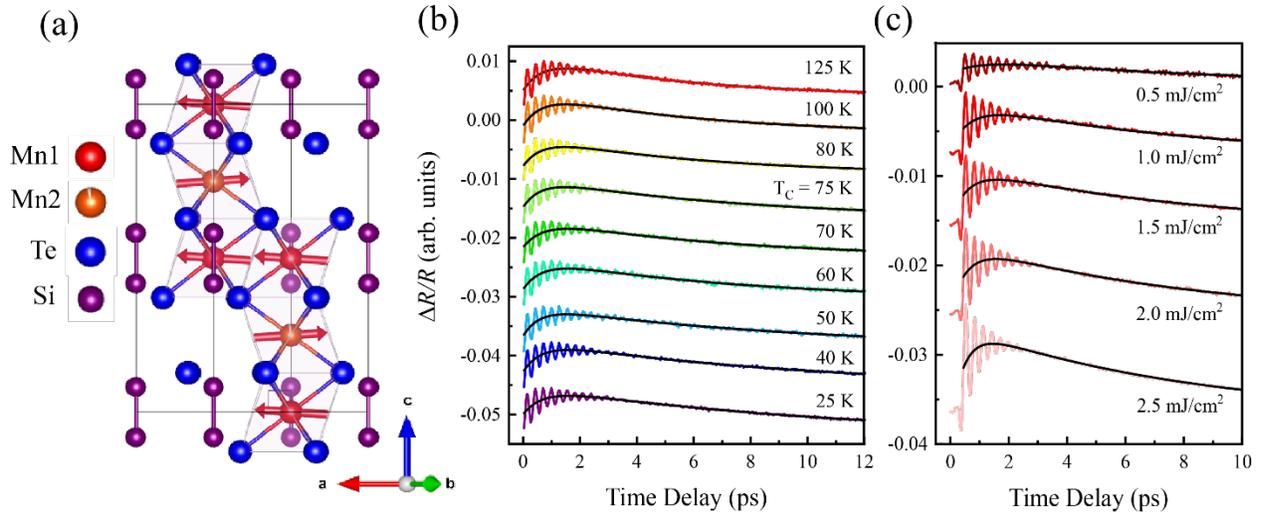

**FIG 1.** (a) The self-intercalated trigonal structure of Mn$_3$Si$_2$Te$_6$ with its magnetic moments (red arrows) lying within the *ab*-plane, (b) $\Delta R/R$ from 25 - 125 K at a pump fluence of 1.5 mJ/cm$^2$, and (c) the $\Delta R/R$ traces with varying pump fluence measured at 300 K. The black lines in both (b) and (c) are fits to the data using Eq. 1.

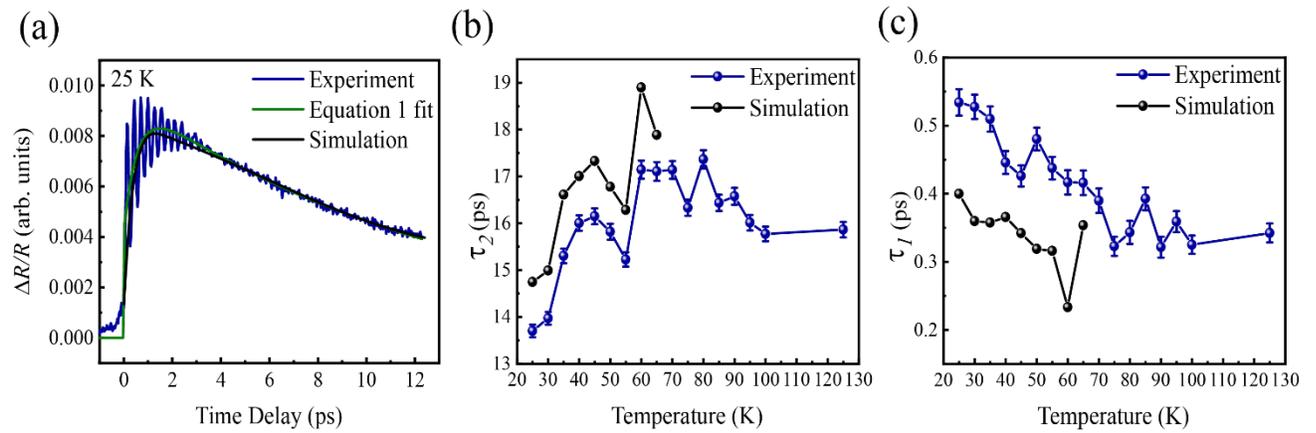

**FIG 2.** (a) The experimental $\Delta R/R$ (blue) measured at 25 K with the biexponential fit from Eq. 1 (green) and the M4TM simulation (black). (b) The temperature dependence of $\tau_1$ (rise time) and (c) $\tau_2$ (decay time) along with the time constants extracted from the M4TM simulation.

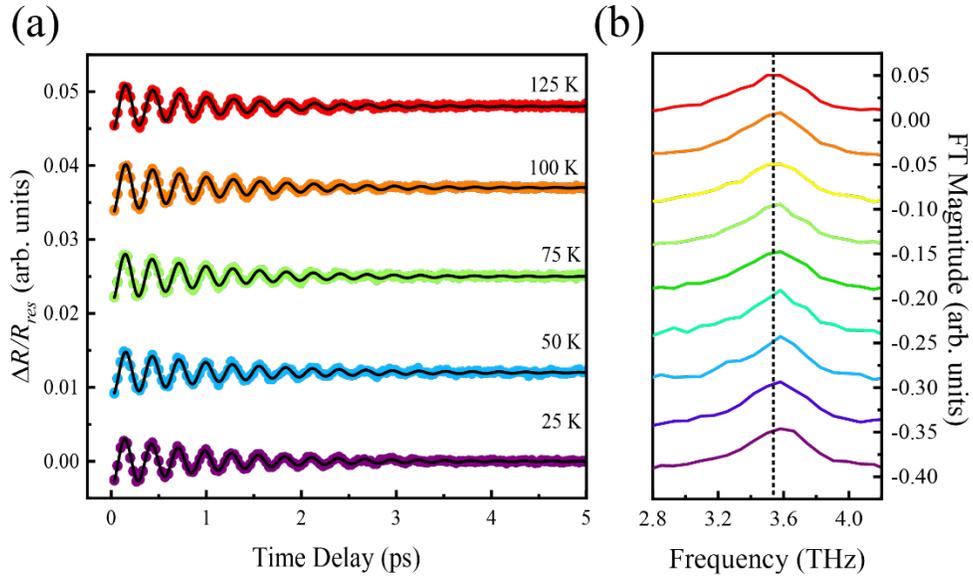

**FIG 3.** (a) The experimental $\Delta R/R_{res}$ at various temperatures above and below $T_c$ and (b) the magnitude of their respective FTs. The black lines in (a) are fits calculated from the damped sine function (Eq. 5) and the dotted line in (b) denotes the high temperature frequency of the phonon mode (3.52 THz).

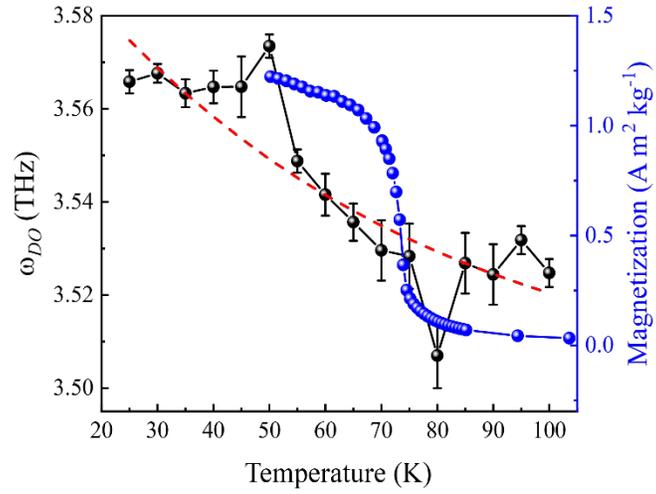

**FIG 4.** The damped oscillation frequency (black dots, left y-axis) obtained from Eq. 5, plotted as a function of temperature and the temperature dependent magnetization (blue dots, right y-axis) measured with an applied field of 0.5 T. The red dashed line is the expected temperature variation of the phonon frequency under a thermal expansion model (Eq. 6) with $\gamma = -5.271$, $\Gamma_{L,0} = 0.00413$ THz, $\omega_0 = 3.607$ THz, and $\lambda_{ph-ph} = 0.6$.

# Supplemental Material for "Coherent Spin-Phonon Coupling in the Layered Ferrimagnet $Mn_3Si_2Te_6$"

L. M. Martinez[1,2], Y. Liu[3,¶], C. Petrovic[3], S. Haldar[4], T. Griepe[5], U. Atxitia[5],

M. Campbell[1,6], M. Pettes[1], R. P. Prasankumar[1],

E. J. G. Santos[4,7*], S. R. Singamaneni[2,‡], and P. Padmanabhan[1,†]

[1]Center for Integrated Nanotechnologies, Los Alamos National Laboratory, Los Alamos, NM 87545

[2]The University of Texas at El Paso, Department of Physics, El Paso, TX 79968

[3]Brookhaven National Laboratory, Exploratory Materials Synthesis & Characterization, Upton, NY 11973

[4]Institute for Condensed Matter Physics and Complex Systems, School of Physics and Astronomy, University of Edinburgh, Edinburgh EH9 3FD, United Kingdom

[5]Instituto de Ciencia de Materiales de Madrid, CSIC, Cantoblanco, 28049 Madrid, Spain

[6]University of California Irvine, Department of Physics & Astronomy, Irvine, CA 92697

[7]Higgs Centre for Theoretical Physics, The University of Edinburgh, EH9 3FD, United Kingdom

†prashpad@lanl.gov

*esantos@ed.ac.uk

‡srao@utep.edu

¶Present address: Los Alamos National Laboratory, Los Alamos, New Mexico 87545, USA

Below, we show additional experimental parameters obtained from Equation 1 in the main text, the extended time scans measured as a function of pump fluence, and the theoretical methods, simulations of the experimental data and parameters obtained from the microscopic four temperature model (M4TM).

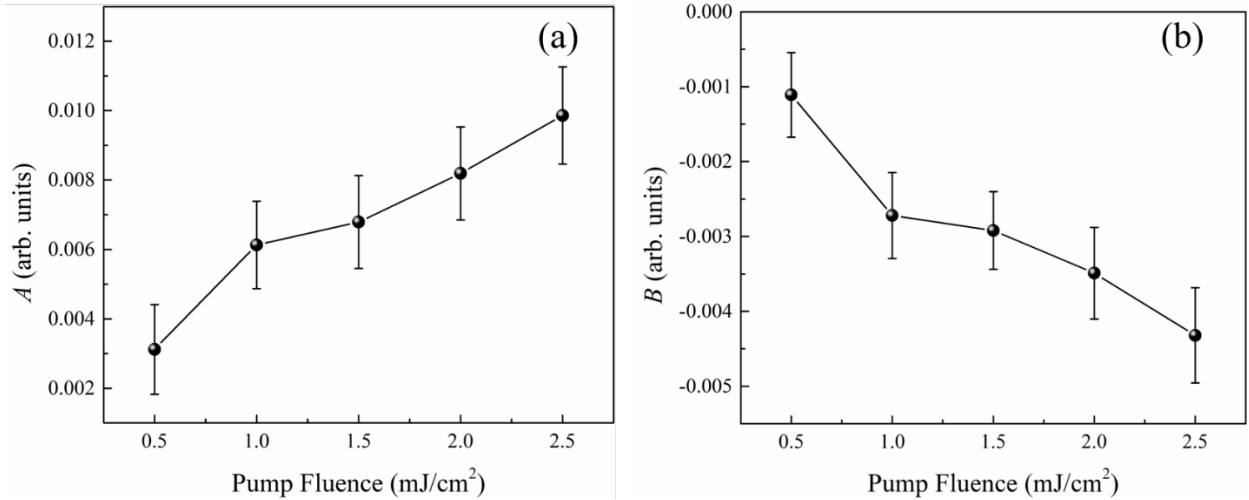

Supp. Mat. Figure 1: The (a) A and (b) B amplitudes obtained from the biexponential model (Eq. 1 in the main text). We observe a linear increase (decrease) in the amplitude of $A$ ($B$) as the pump fluence increases.

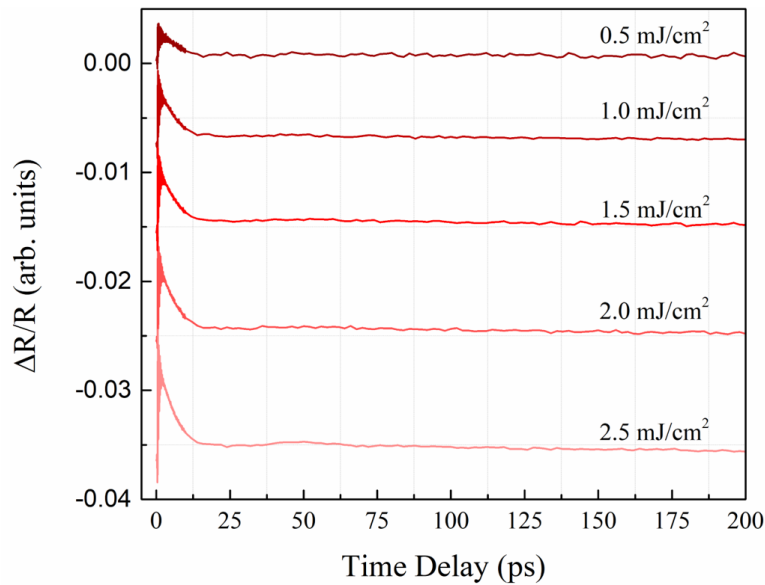

Supp. Mat. Figure 2: The differential reflectivity ($\Delta R/R$) of $Mn_3Si_2Te_6$ (MST) measured at extended time delays obtained at 300 K at various pump fluences. No significant dynamics were observed at longer time scales.

# Microscopic Four Temperature Model

In the main text we go over some the initial aspects of the M4TM, however, there are additional equations that go into this model to work within our material system MST.

The magnetization ($m$) dynamics are described in conjunction with the change in the occupation ($f_{m_s}$) of the z-component of $m_s$,

$$\frac{dm}{dt} = -\frac{1}{S}\sum_{m_s=-S}^{m_s=+S} m_s \frac{df_{m_s}}{dt} \tag{SM1}$$

$$\frac{df_{m_s}}{dt} = -(W^+_{m_s} + W^-_{m_s})f_{m_s} + W^+_{m_s-1}f_{m_s-1} + W^-_{m_s+1}f_{m_s+1} \tag{SM2}$$

$$W^{\pm}_{m_s} = R\frac{Jm}{4Sk_BT_C}\frac{T_{po}}{T_C}\frac{e^{\pm\frac{Jm}{2Sk_BT_e}}}{\sinh\left(\frac{Jm}{2Sk_BT_e}\right)}\left(S(S+1) - m_s(m_s \pm 1)\right). \tag{SM3}$$

The rate parameter $R$ depends on the material and is proportional to the spin-flip probability, $a_{sf}$. This model is true for arbitrary spin $> \frac{1}{2}$, whereas the microscopic three-temperature model [1] is valid only for spin-1/2. The first term signifies the reduction of the occupation numbers upon scattering events at either higher or lower spin levels, and the second terms correspond to the increase of occupation numbers at those levels. Beens *et al.* [2] first calculated the transition rate $W^{\pm}_{m_s}$ from the Fermi's golden rule as

$$H_{eps} = \sqrt{\frac{a_{sf}}{D_s}\frac{\lambda_{ep}}{N^{3/2}}}\sum_{k,k'}\sum_{q}^{ND_p}\sum_{j}^{N_s} c^{\dagger}_k c_{k'}(S_{j,+} + S_{j,-})(a^{\dagger}_q + a_q) \tag{SM4}$$

The rate parameter in equation (SM3) is then,

$$R = 8\frac{a_{sf}g_{e-po}T_C^2V_{at}}{\mu_{at}k_BT_{Debye}^2} \tag{SM5}$$

Where $\mu_{at}$ is the atomic magnetic moment in the units of Bohr magneton and $k_B$ is the Boltzmann constant.

The temporal shape of the pump pulse can be approximately described with a Gaussian function,

$$P(t) = P_0 exp\left\{-\frac{(t-t_0)^2}{2\sigma^2}\right\}, \tag{SM6}$$

centered around $t_0$ and pulse width $\sigma$. $P_0$ is the energy density of the laser pulse absorbed by the electron system. The heat capacity of the electron system is calculated in the Sommerfeld approximation as

$$C_e = \gamma_e T_e \tag{SM7}$$

where $\gamma_e$ is the Sommerfeld coefficient, and the heat capacity of the phonon is computed with the Einstein model,

$$C_{po} = C_{po\infty} \frac{T_{Ein}^2}{T_{po}^2} \frac{\exp\{\frac{T_{Ein}}{T_{po}}\}}{\left(\exp\{\frac{T_{Ein}}{T_{po}}\}-1\right)^2} \quad (SM8)$$

Debye temperature ($T_{Debye}$) related to the Einstein temperature $T_{Ein}$ through the relation $T_{Ein} = 0.75 T_{Debye}$. Therefore, we use experimentally measured Debye temperature as input for our model in the intermediate temperature region.

The accurate value of $g_{e-po}$ is essential for the description of the magnetization dynamics in the initial phase. However, the experimental value of $g_{e-po}$ for MST is unavailable in the literature, we fit the experimental data to get the parameter.

## Characterization of the pump pulse

In our M4TM simulations, we take the pump pulse width as $\sigma = 60$ fs and fluence as $F = 1.5$ mJ/cm$^2$. The absorbed pump power density $P_0$ in the monolayer from the incident fluence is given by,

$$P_0 = \frac{F}{\sqrt{2\pi} P_0 \sigma 10} \quad (SM9)$$

where the unit of $P_0$ is $\frac{W}{m^3}$. We further assume a homogenous heat distribution over the sample thickness by the laser pulse.

## Simulation setup; input parameters

Supp. Mat. Table 2 shows the parameters we keep fixed for the simulations, and the parameters in Supp. Mat. Table 3 were varied for different temperatures.

Supp. Mat. Table 1 shows the fixed input parameters used for the M4TM.

| Symbol | Description | Value | Units |
|---|---|---|---|
| $g_{po-pa}$ | Ph-ph coupling | $1.82 \times 10^{18}$ | $W/m^3 K$ |
| $T_C$ | Curie temp. | 75[3] | K |
| $T_{Debye}$ | Debye temp. | 155[3] | K |
| $T_{Ein}$ | Einstein temp. | 116 | K |
| S | Effective spin | 5/2[3] | |
| $\mu_{at}$ | Atom. Mag. moment | 1.6 | $\mu_B$ |
| $V_{at}$ | Atomic volume | 117 | Å$^3$ |

Supp. Mat. Table 2 shows the variable parameters extracted from the M4TM. These parameters are the electron-phonon coupling constant ($g_{e-po}$), Sommerfield coefficient ($\gamma_e$), heat capacities for the optical ($C_{po}$) and acoustic ($C_{pa}$) phonons subsystems and the spin-flip probability ($a_{sf}$).

| T (K) | $g_{e-po}$ (W/m³K) | $\gamma_e$ (J/m³K²) | $C_{po}$ (J/m³K) | $C_{pa}$ (J/m³K) | $a_{sf}$ |
|---|---|---|---|---|---|
| 25 | $1.75 \times 10^{18}$ | $4.4 \times 10^3$ | $3.2 \times 10^7$ | $3.0 \times 10^6$ | 0.180 |
| 30 | $1.85 \times 10^{18}$ | $4.0 \times 10^3$ | $4.0 \times 10^7$ | $2.0 \times 10^6$ | 0.150 |
| 35 | $1.85 \times 10^{18}$ | $4.0 \times 10^3$ | $5.3 \times 10^7$ | $1.0 \times 10^6$ | 0.125 |
| 40 | $1.95 \times 10^{18}$ | $4.4 \times 10^3$ | $7.8 \times 10^7$ | $1.0 \times 10^6$ | 0.115 |
| 45 | $2.05 \times 10^{18}$ | $4.0 \times 10^3$ | $1.6 \times 10^8$ | $1.0 \times 10^6$ | 0.105 |
| 50 | $2.35 \times 10^{18}$ | $4.0 \times 10^3$ | $8.6 \times 10^8$ | $4.0 \times 10^6$ | 0.105 |
| 55 | $3.55 \times 10^{18}$ | $8.0 \times 10^3$ | $9.6 \times 10^8$ | $4.0 \times 10^6$ | 0.100 |
| 60 | $4.80 \times 10^{18}$ | $8.0 \times 10^3$ | $8.6 \times 10^9$ | $5.0 \times 10^6$ | 0.100 |
| 65 | $6.00 \times 10^{18}$ | $13.4 \times 10^3$ | $7.8 \times 10^{10}$ | $7.0 \times 10^6$ | 0.185 |
| 70 | $9.00 \times 10^{18}$ | $19.4 \times 10^3$ | $9.8 \times 10^{10}$ | $8.0 \times 10^6$ | 0.385 |

**Fitting procedure**

Comparing the experimental data, we extracted the value of the electron-phonon coupling, phonon-phonon coupling, Sommerfeld constant, and the heat capacities of optical and acoustic phonons for different temperatures. The other parameters which we have used for the simulation are taken from the literature. First, the electron-phonon coupling parameter was identified as this is responsible for the initial demagnetization phase occurring in the initial 2 ps. After that, Sommerfeld constant, optical phonon heat capacity and spin-flip probability are adjusted by the longer time scale magnetization dynamics from 2ps until magnetization equilibrium is reached.

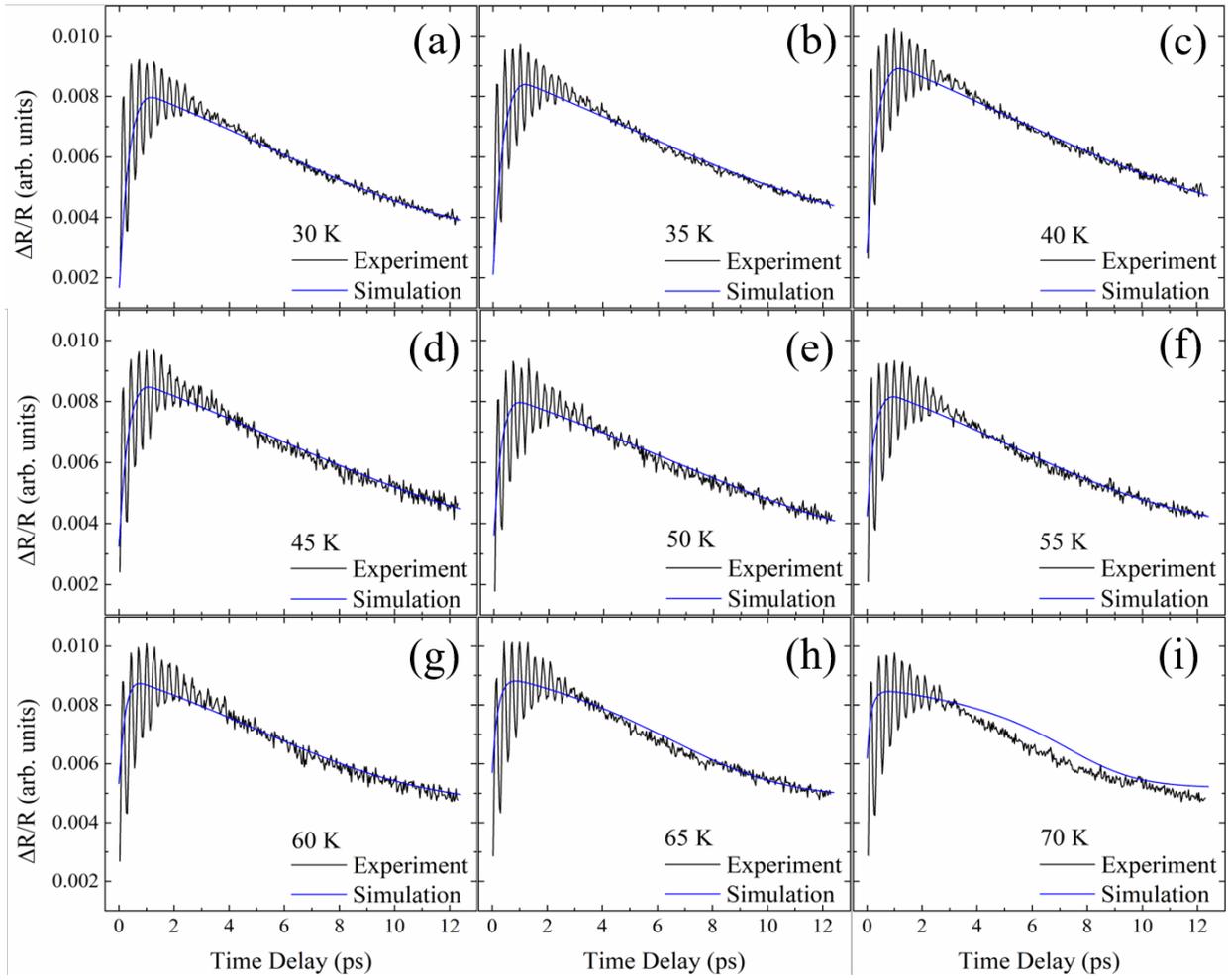

Supp. Mat. Figure 3: The experimental $\Delta R/R$ from (a - i) 30 – 70 K along with the simulated incoherent dynamics from the M4TM. We see that the correspondence between the experimental data and M4TM improve deeper into the magnetic phase of MST.

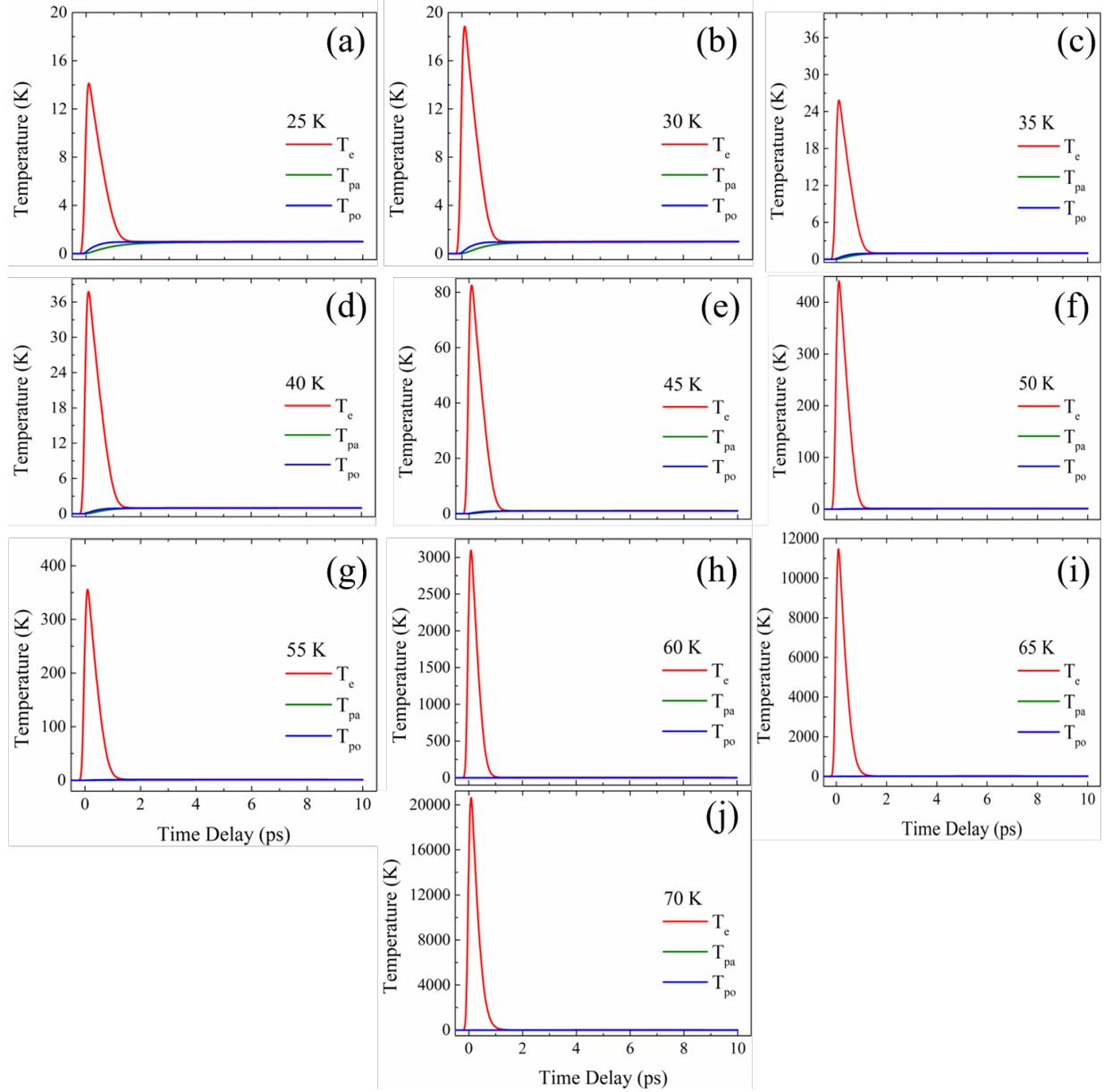

Supp. Mat. Figure 4: The electronic ($T_e$), acoustic ($T_{pa}$) and optical phonon ($T_{po}$) temperatures obtained from the M4TM simulations from (a – j) 25 – 70 K.

**References**

1. B. Koopmans, G. Malinowski, F. Dalla Longa, D. Steiauf, M. Fähnle, T. Roth, M. Cinchetti, and M. Aeschlimann. Explaining the paradoxical diversity of ultrafast laser-induced demagnetization. *Nature Materials 9, 259-265 (2010)*.
2. M. Beens, M. L. M. Lalieu, A. J. M. Deenen, R. A. Duine, and B. Koopmans. Comparing all-optical switching in synthetic ferrimagnetic multilayers and alloys. *Phys. Rev. B 100, 220409(R) (2019)*.
3. Y. Liu, Z. Hu, M. Abeykoon, E. Stavitski, K. Attenkofer, E. D. Bauer, and C. Petrovic. Polaronic transport and thermoelectric in $Mn_3Si_2Te_6$ single crystals. *Phys. Rev. B 103, 245122 (2021)*.